\documentstyle[aps,prd,eqsecnum]{revtex}
\input psfig.sty

\def\be{\begin{equation}}
\def\ee{\end{equation}}
\def\ba{\begin{eqnarray}}
\def\ea{\end{eqnarray}}

\begin{document}

\draft

\title{Gravity Waves, Chaos, and Spinning Compact Binaries}

\author{Janna Levin}
\address{DAMTP, Cambridge University,
Wilberforce Rd., Cambridge CB3 0WA }
%\quad J.Levin@damtp.cam.ac.uk}
\address{and Astronomy Centre, University of Sussex,Brighton BN1
9QJ}

\twocolumn[\hsize\textwidth\columnwidth\hsize\csname
           @twocolumnfalse\endcsname

\maketitle
\widetext

\begin{abstract}

Spinning compact binaries are shown to be chaotic in the
Post-Newtonian expansion of the two body system.  
Chaos by definition is the extreme sensitivity to
initial conditions and a consequent inability to predict the outcome
of the evolution.  
As a result, the spinning pair will have
unpredictable gravitational waveforms during coalescence.  
This poses a challenge to future gravity wave
observatories which rely on a match between the data and a theoretical
template. 

\end{abstract}

\medskip
\noindent{04.30.Db,97.60.Lf,97.60.Jd,95.30.Sf,04.70.Bw,05.45}
\medskip
]

\narrowtext
 
\setcounter{section}{1}

Coalescing
binaries are the primary objects of attention for future ground based
gravity wave detectors such as LIGO and VIRGO.  The successful
detection of the waveforms requires a technique of matched filtering
whereby the data is convolved with a theoretical template.  Excellent
agreement is required if a signal is to be drawn out of the noise.
A possible obstacle to the method of matched filtering can surface if
the orbits become chaotic.
As shown here, the final coalescence of spinning,
compact binaries proceeds chaotically for some spin configurations.
Chaotic binaries with similar initial conditions 
may produce disparate waveforms and consequently they
may not be detectable by the method of matched
filtering.
An alternative method must be sought for their detection.  

Many authors have emphasized that black holes are susceptible to 
chaos \cite{{bhs},{bc},{maeda},{carl},{cf},{me}}.  
Chaos has not received
the attention it deserves in part because the systems studied have
been highly idealized.  An elegant example of chaos around
black holes is provided by the Majumdar-Papapetrou spacetimes
\cite{{maj},{pap}}
which
arrange extremal black holes such that the gravitational attraction
of their masses is exactly countered by the electrostatic repulsion
of their charges.  The spacetime is static and yields a simple
solution.  The geodesics however are formally non-integrable and fully
chaotic \cite{{bhs},{carl}}.
A static spacetime produces no gravitational waves and so the chaotic
scattering in the Majumdar-Papapetrou spacetime remains just an
interesting formal system, although gravity waves are produced by a third 
orbiting body \cite{cf}.  
Chaos around Schwarzschild black holes has also been studied formally
with a hypothetical perturbation of a test companion along
the homoclinic orbits which mark the boundary between
dynamical stability and instability
\cite{bc}.  
Another important example
of chaos around a black
hole is the motion of a spinning test particle \cite{maeda}.  
This already shows the key features of the two-body system investigated here.

In this paper, the most realistic description currently available
of a black hole plus a
companion is shown to succumb to chaos when the
pairs spin.  The Post-Newtonian (PN) expansion of the relativistic
two-body problem \cite{{pn},{dd},{ww},{iyer}} 
provides the dynamical equations of motion
to 2PN-order \cite{{kidder},{kww2}}.  
In the absence of spins, the existence of a conserved
angular momentum and energy \cite{dd} 
ensure that the system is in principle integrable to at least
5/2PN-order
\cite{us}.  The non-spinning pair still has two identifiable circular 
orbits for a given angular momentum, one stable
and one unstable.  
In the transition to chaos, the periodic orbits proliferate
and these form the structure of the chaotic dynamics.
The homoclinic orbits found in Ref. \cite{us} 
demarcate the region of phase space at which this occurs, perhaps
at higher orders in the PN expansion.
%(The full relativistic two-body problem gives every indication of being
%non-integrable and hence fully chaotic.)

When spins are introduced at 2PN-order, 
the orbital plane precesses chaotically.
There are now an infinite number of periodic orbits which form a
fractal in the dynamical phase space.  We can isolate this fractal
through the method of fractal basin boundaries
\cite{{carl},{cf},{me},{ott},{nj},{mixm}}.  
Fractals are a particularly important tool
in relativity since they do not depend on the coordinate system used,
a point emphasized in \cite{mixm}.

In the notation of Ref. \cite{kidder}, the center of mass equations of
motion in harmonic coordinates are
	\be
	\ddot{ \vec x }=\vec a_{PN} +\vec a_{SO}+\vec
	a_{SS}+\vec a_{RR} .\label{eom}
	\ee
The right hand side is the sum of the contributions to the relative
acceleration from the PN expansion, from the spin-orbit (SO)
and
spin-spin (SS) coupling and from the 
radiative reaction (RR).
The spins also precess by
	\be
	\dot{ \vec S_1} = \vec \Omega_1\times \vec S_1 \quad  , \quad
	\dot{ \vec S_2 }= \vec \Omega_2\times \vec S_2 .\label{prec}\ee
 For brevity we do not rewrite the explicit
forms of $\vec a$ and $\vec \Omega$ here but they can be found in
Ref. \cite{kidder}.
There are 12 degrees of freedom $(\vec x,\dot {\vec x}, \vec S_1, \vec
S_2)$.  The form of eqn.\ (\ref{prec}) indicates that the magnitudes of
the individual spins are conserved.  To 2PN-order there is also a
conserved 
energy $E$ and a conserved
total angular momentum $\vec J=\vec L+\vec S$ where $\vec L$ is the
orbital angular momentum and $\vec S=\vec S_1+\vec S_2$.  In all, there
are $6$ constants of motion reducing the phase space to $6$ degrees of
freedom, plenty to allow for chaotic motion.
The condition that the orbit be perfectly circular $\dot r=\ddot r=0$
(where $r=|\vec x|$)
still leads to an underdetermined set of equations for which there are
an infinite number of spin configurations.  
This is evidence for the proliferation of periodic orbits and
indicates the pursuit of an
innermost stable circular orbit \cite{{rdet}} is futile.

\begin{figure}
\centerline{\psfig{file=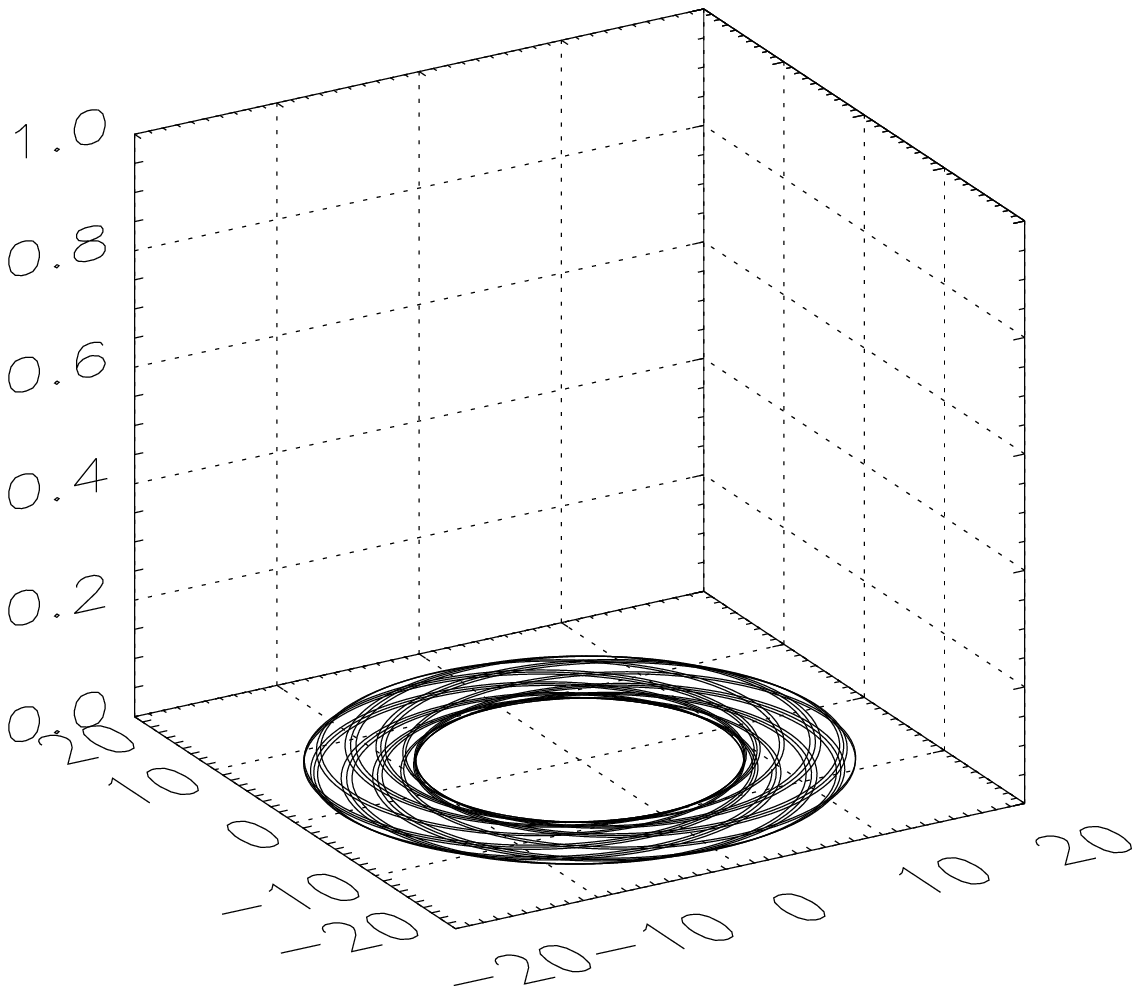,width=3.in}}
\centerline{\psfig{file=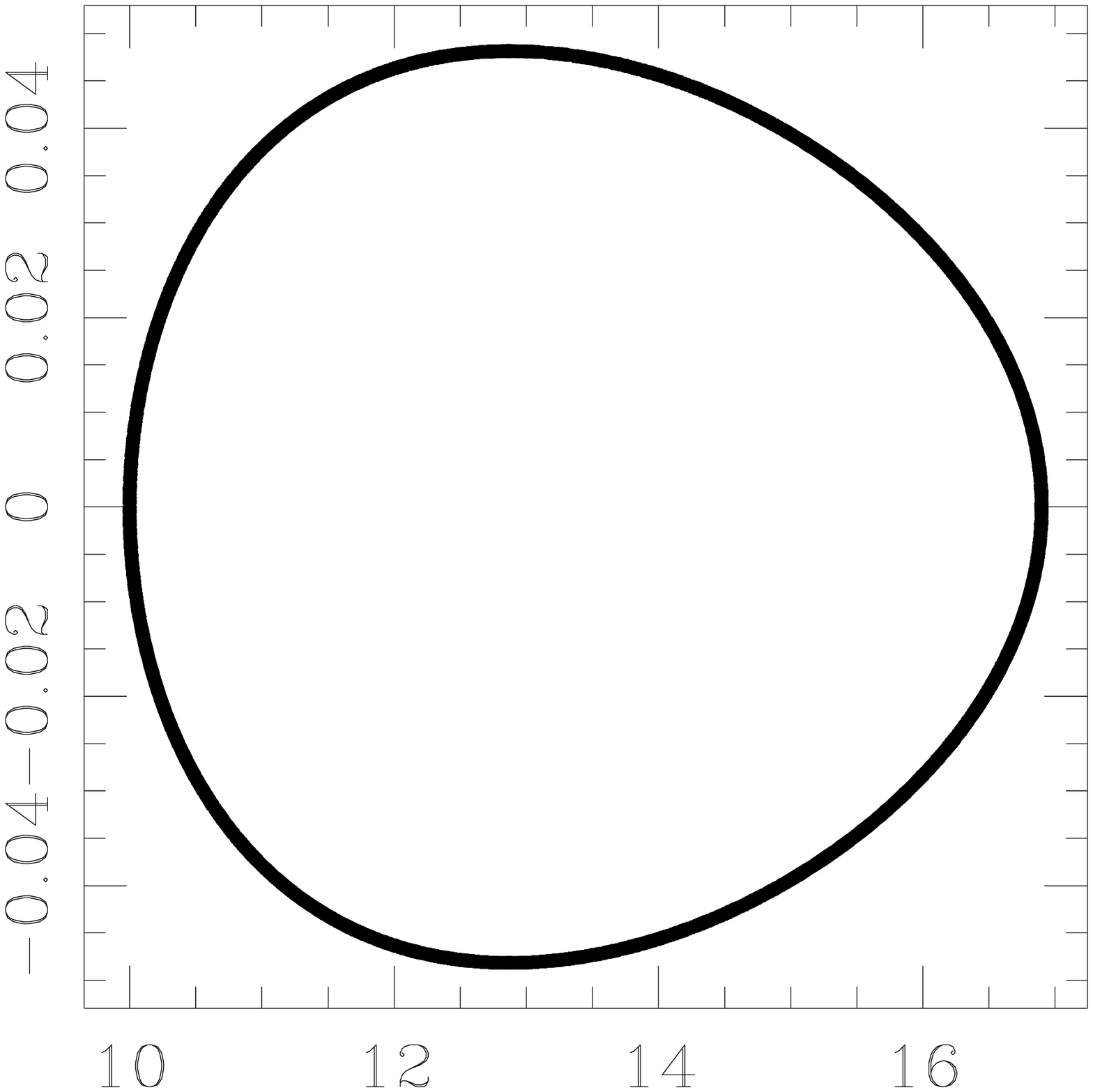,width=1.85in}
{\psfig{file=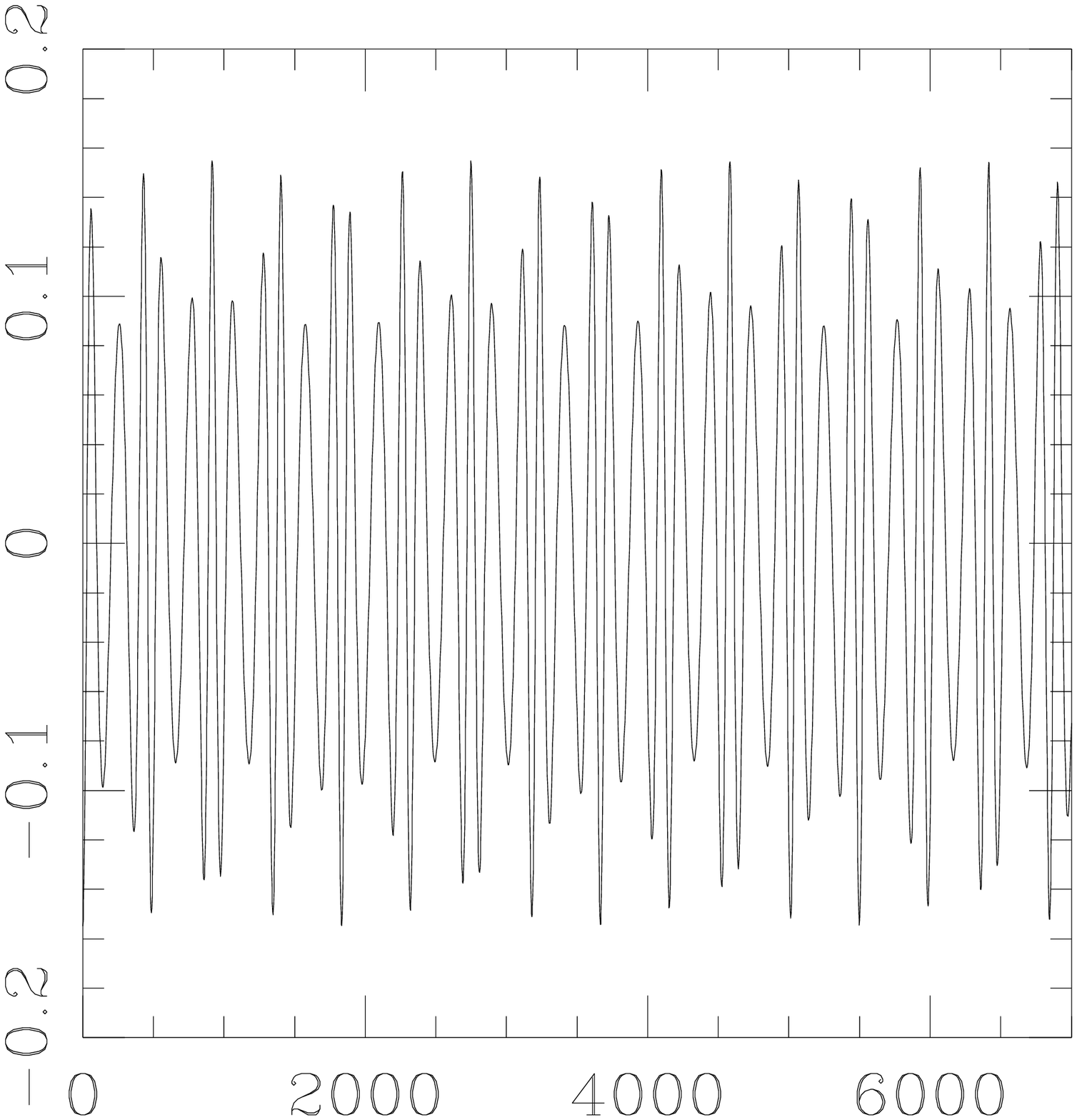,width=1.85in}}}
\caption{The pair has mass ratio $m_2/m_1=1.4/10$ and no spins.
The
initial conditions are 
$x_i/m=10, \dot y_i=0.3 $ and $z_i=0$.  Time is measured in units
of the total mass $m$.
Top: A 3D view of the orbit.  Lower Left: The smooth phase space curve
in the $(r,\dot r)$ plane.
Lower Right:  The waveform $h_{+}$.
\label{orb0}}  \end{figure}

Figure \ref{orb0} shows typical orbital motion in the absence of
spins and with the dissipative (RR)-term in eqn.\ (\ref{eom}) temporarily
turned off.   There is no precession of the orbital plane and no chaos.  
Although the orbit is confined to a plane, the perihelion precesses within
the  plane due to the relativistic corrections.  
The regularity of the motion is confirmed by the phase space diagram
in fig.\ \ref{orb0}
which shows the motion to be confined to a smooth line in the
$(r,\dot r)$ plane.  
The waveforms for specific orbits are
obtained to 3/2PN-order using the results of Ref. \cite{kidder} and
neglecting tail contributions.
For simplicity we show the $+$-polarization waveform, 
$h_+=h_{xx}$, with the Earth located above the $z$-axis.
%The waveform in the bottom panel of fig.\ \ref{orb0} is reminiscent of
%the waveforms for relativistic orbits found in Ref. \cite{us}.

\begin{figure}
\centerline{\psfig{file=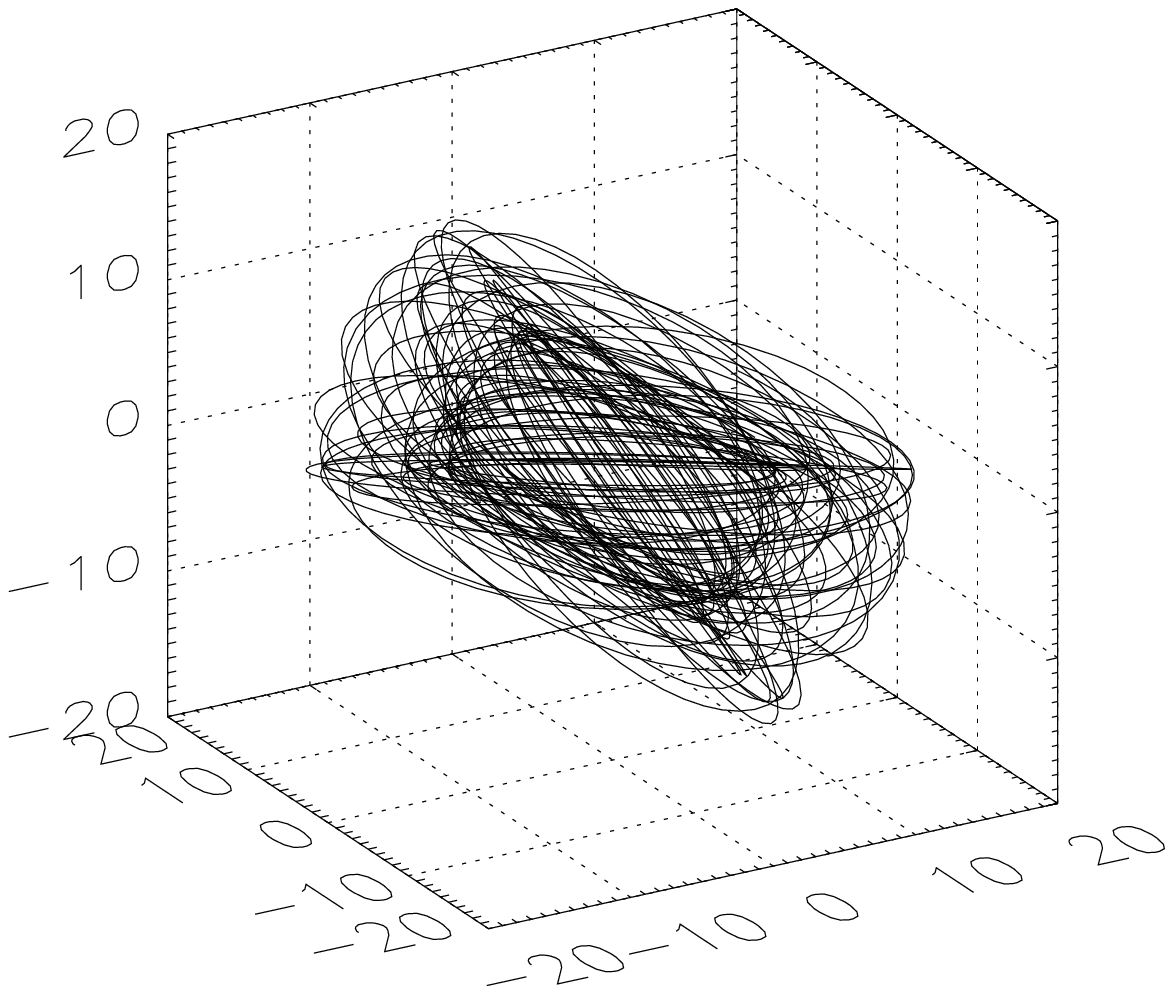,width=3.in}}
\centerline{\psfig{file=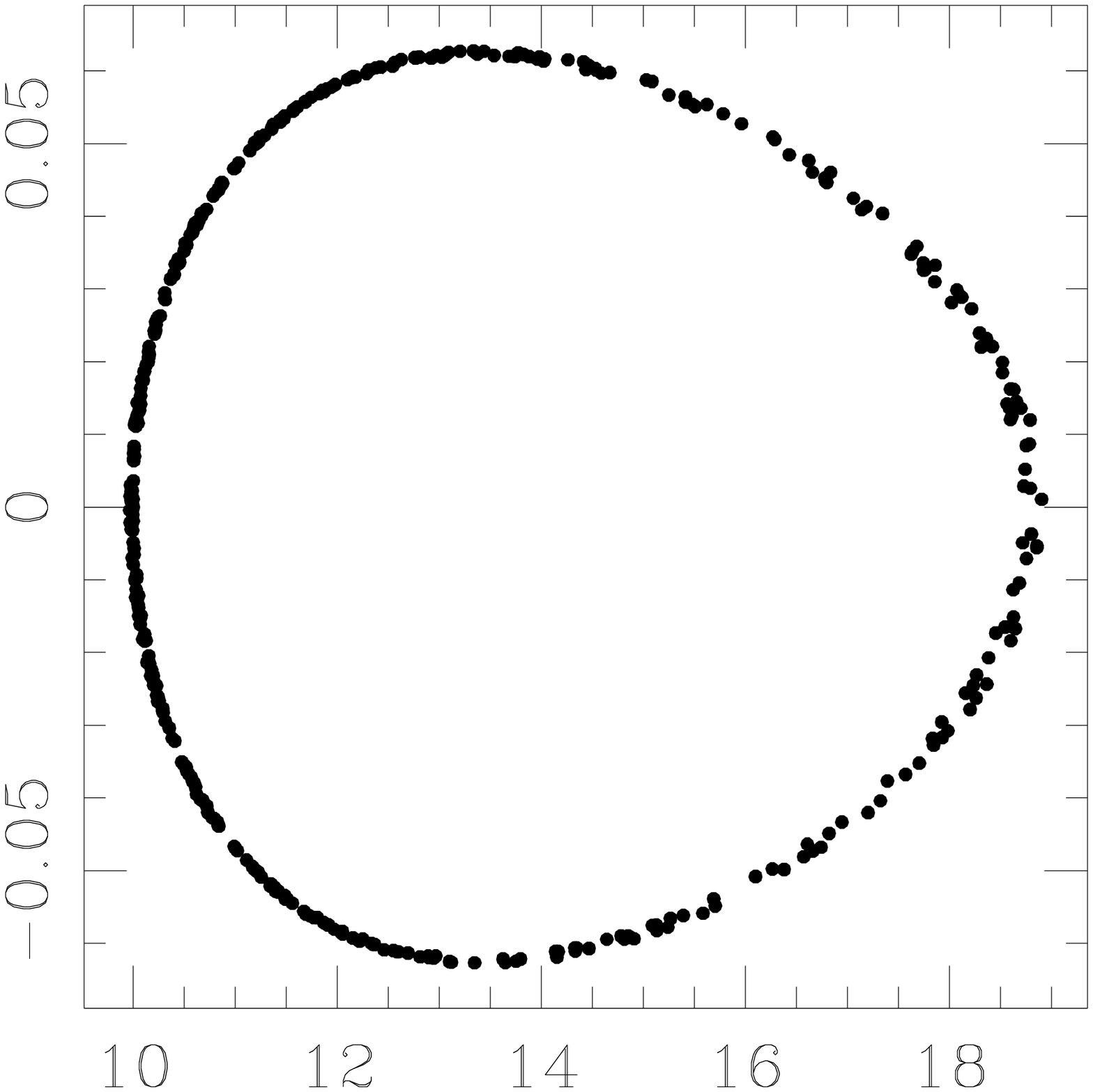,width=1.75in}
\psfig{file=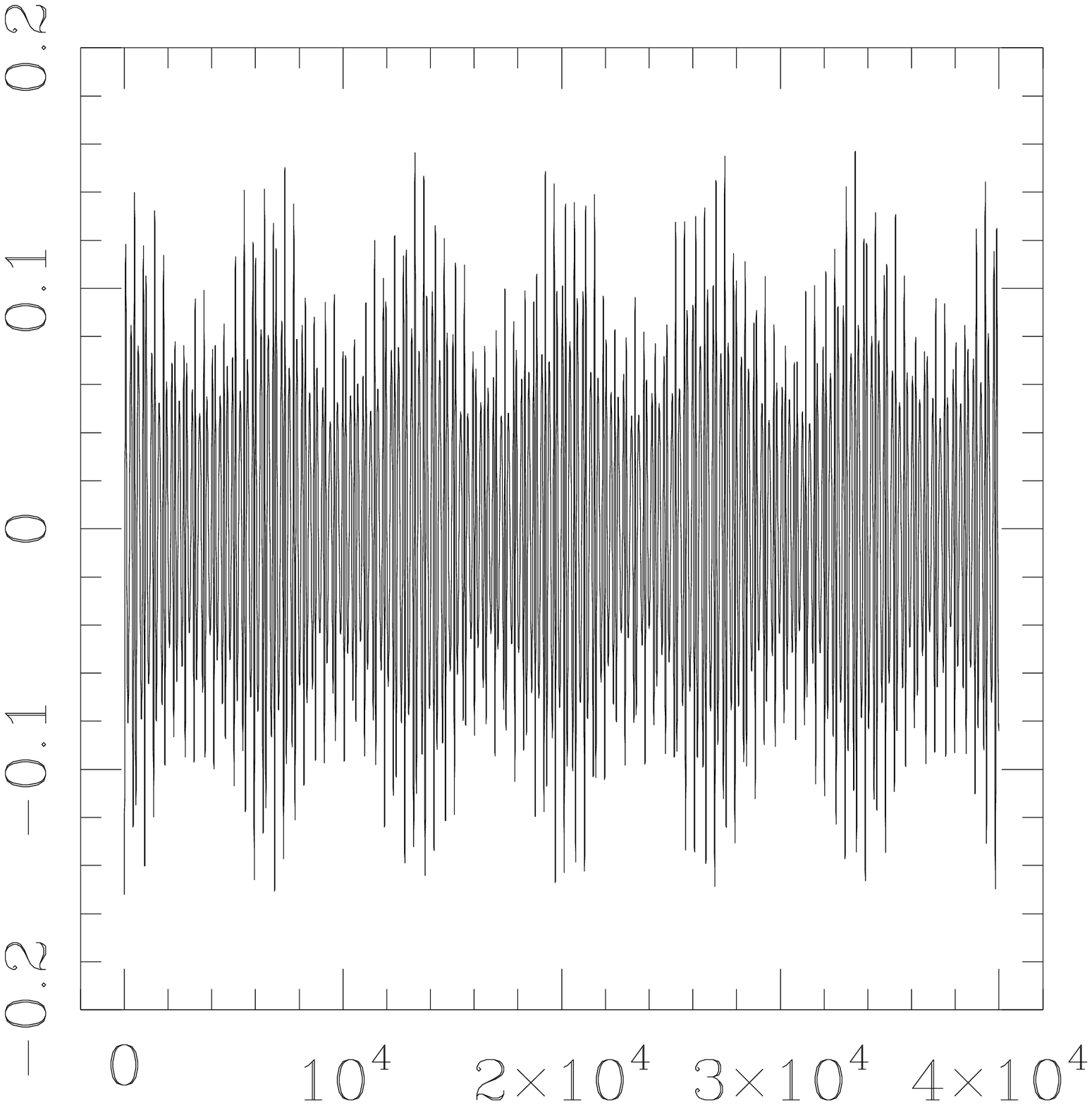,width=1.75in}}
\caption{
The pair has mass ratio $m_2/m_1=1.4/10$ and spins
$S_1= m_1^2, S_2=0.7m_2^2$.  
The
initial conditions are 
$x_i/m=10, \dot y_i=0.3 $ and $z_i=0$.  
The initial angles are $\theta_1=\theta_2=45{}^o$. 
Top: A 3D view of the orbit.
Lower Left: The surface of section
in the $(r,\dot r)$ plane.
Lower Right: The waveform $h_+$.
\label{orb4}}  \end{figure}

If the compact objects spin,
then the motion can become chaotic.
The spin vector $\vec S_1$ is tilted
by an angle $\theta_1$ measured from
the $\hat z$-axis and the spin vector $\vec S_2$ is tilted by an angle
$\theta_2$.  
The motion is clearly occupying three dimensions
and is no longer confined to a plane as demonstrated in fig.\ \ref{orb4}.
A Poincar\'e surface of section is constructed by plotting a point
as the orbit crosses the $z=0$ plane from $z>0$ to 
$z<0$.
A regular orbit would draw a smooth curve in the plane while a chaotic
orbit speckles the plane with points unpredictably.
The chaotic precession is indicated in the surface of section which has
begun to turn to dust. 
The more titled the spin vectors, the thicker
the dusty region in the surface of section.
(Due to the large dimensionality of the phase space,
the diagram is a projection onto the
$(r,\dot r)$ plane.
Cautious of any ambiguity this may introduce,
we take the speckled surface only as confirmation of chaos
seen in the precessional motion and
the fractal basin boundaries discussed below.)
The waveform is also shown.

The binary of figure 2 could be a maximally spinning
$10 M_\odot$ black hole with a rapidly rotating $1.4 M_\odot $
neutron star companion.  The spins are each displaced from the initial
orbital angular momentum by $45^o$.
Large spin misalignments 
occur naturally in the formation of
close black hole/neutron star pairs \cite{vk}. 
The orbit shown is within
the LIGO bandwidth with a frequency of roughly ${\cal O}(10-10^2)$ Hz.
With dissipation included, an orbit which begins regular at larger radii
chaotically scatters as the pair draws closer and the signal sweeps
through the LIGO bandwidth.

\begin{figure}
\centerline{\psfig{file=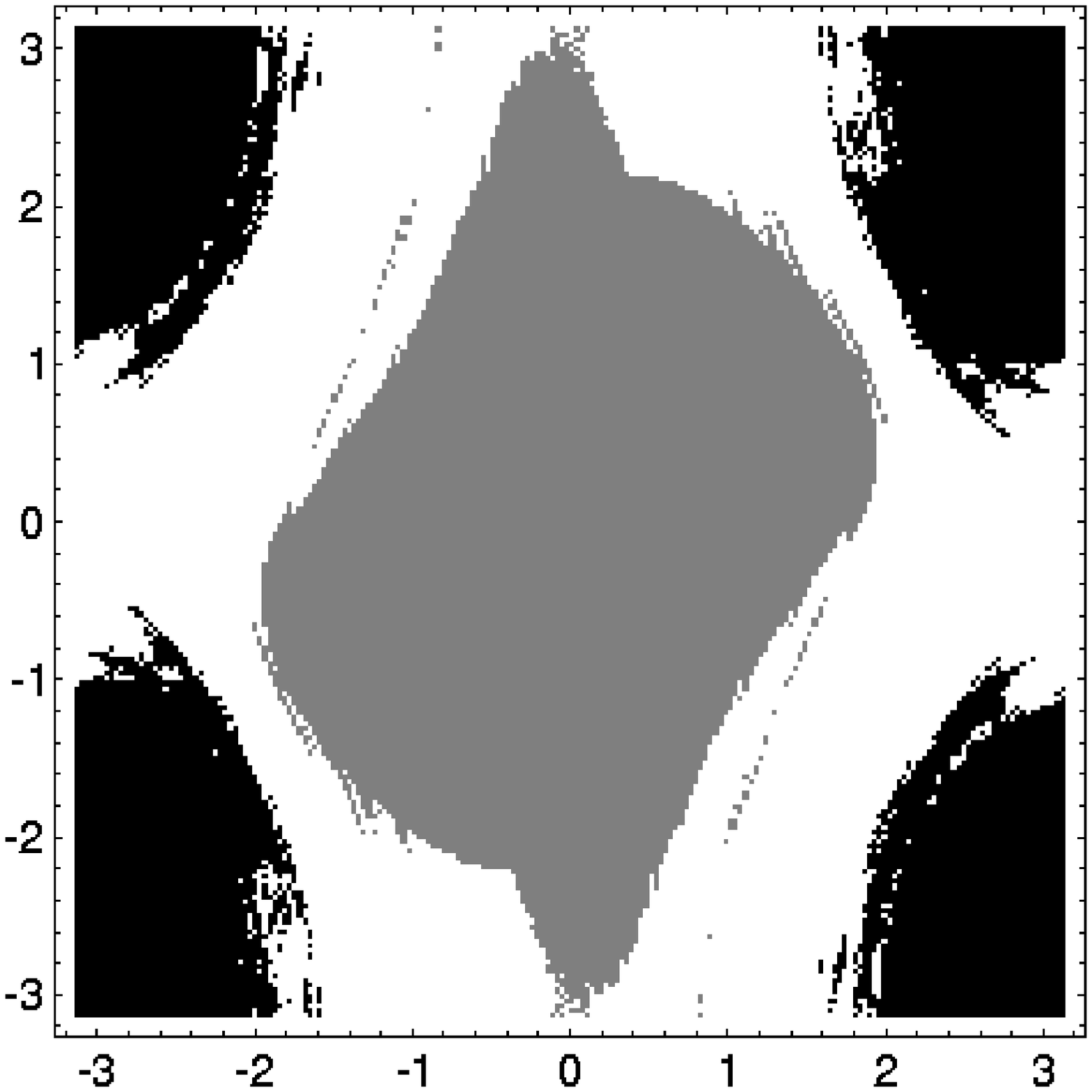,width=2.25in}}
\centerline{\psfig{file=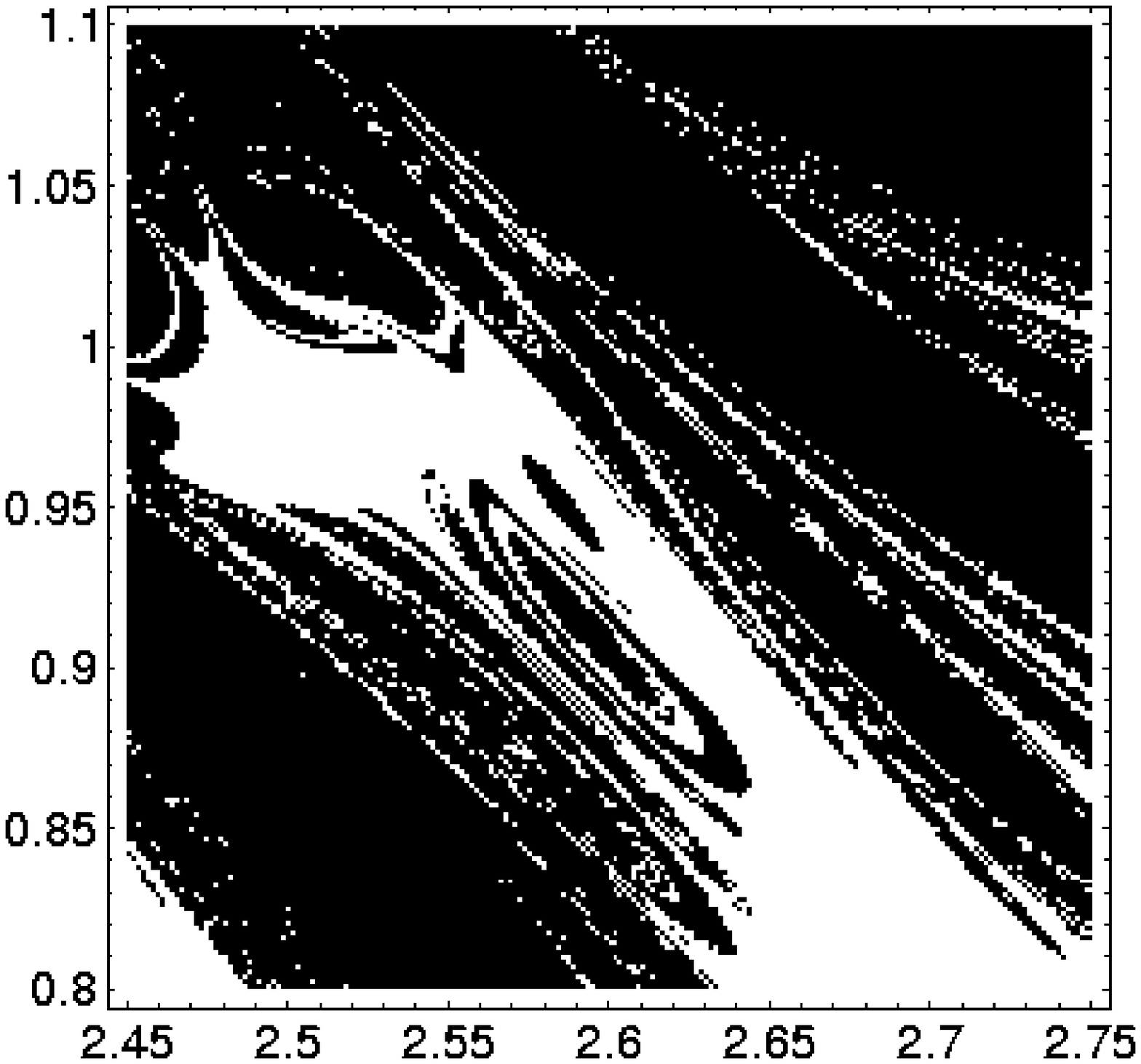,width=2.35in}}
\centerline{\psfig{file=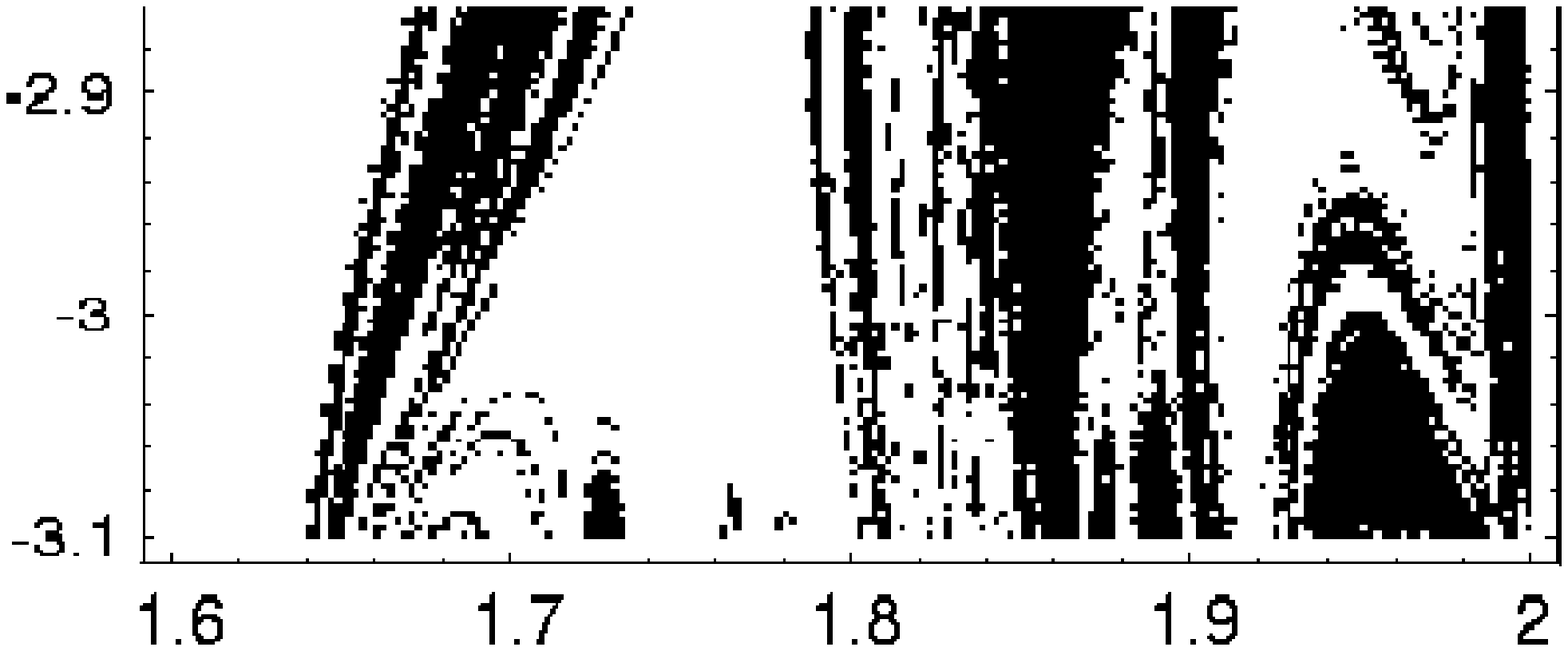,width=2.3in}}
\caption{Top: The fractal basin boundaries for pairs with $m_2/m_1=1/3$ and 
$S_1/m_1^2=S_2/m_2^2=0.6$.  All orbits begin with 
$x_i/m=5 , \dot y_i=0.45$.  
The initial angles $(\theta_1,\theta_2)$ are varied.
The axes are labelled in radians.  $200\times 200$ orbits shown.  
The middle and bottom panels are details of the
upper panel.
\label{fbb}}  \end{figure}

Chaos is not isolated to this specific binary.  Instead of
investigating
individual orbits, we can broadly scan the phase space for chaos.
There may be a sensitivity to the variation of any of the degrees of
freedom as well as the relative masses of the compact objects.  
Since it is impossible to cover all variations, in this instance we
limit
our scan to search for chaos as the spin angles are varied.  To do
this,
we look at a slice through the phase space which varies only the
initial angle $\theta_1$ of $\vec S_1$
and the initial angle $\theta_2$ of $\vec S_2$
for pairs which are otherwise given identical initial conditions
(in this case $m_2/m_1=1/3$ and $S_1/m_1^2=S_2/m_2^2=0.6$).  
These could be black hole pairs.  While the spins are consistent
with neutron star pairs also, they are at such a close separation
that tidal effects for the extended objects would be significant.
The initial location in the $(\theta_1,\theta_2)$ plane is color coded
black if the pair coalesce, grey if the pair separate by $r/m>1000$, 
and white if 
stable motion is attained with more than 50 orbits.
A few pairs which separate to $r/m>1000$ may
still continue orbiting.
Increasing the cutoff would reduce the grey basin.
Also, pushing the stable orbit condition to more than 100 orbits 
tends to
increase the size of the black basins slightly as more orbits have a
chance to coalesce.
If there were no chaos, the boundaries between colors would be smooth
while fractal boundaries signal chaos.  
The fractal basin boundaries of fig.\ \ref{fbb} clearly show a mingling
of possible outcomes as the angles are varied.
The extreme sensitivity to initial conditions is
exemplified in the blown up regions in the lower panels
of fig.\ \ref{fbb} which show the repeated fractal structure.

\begin{figure}
\centerline{\psfig{file=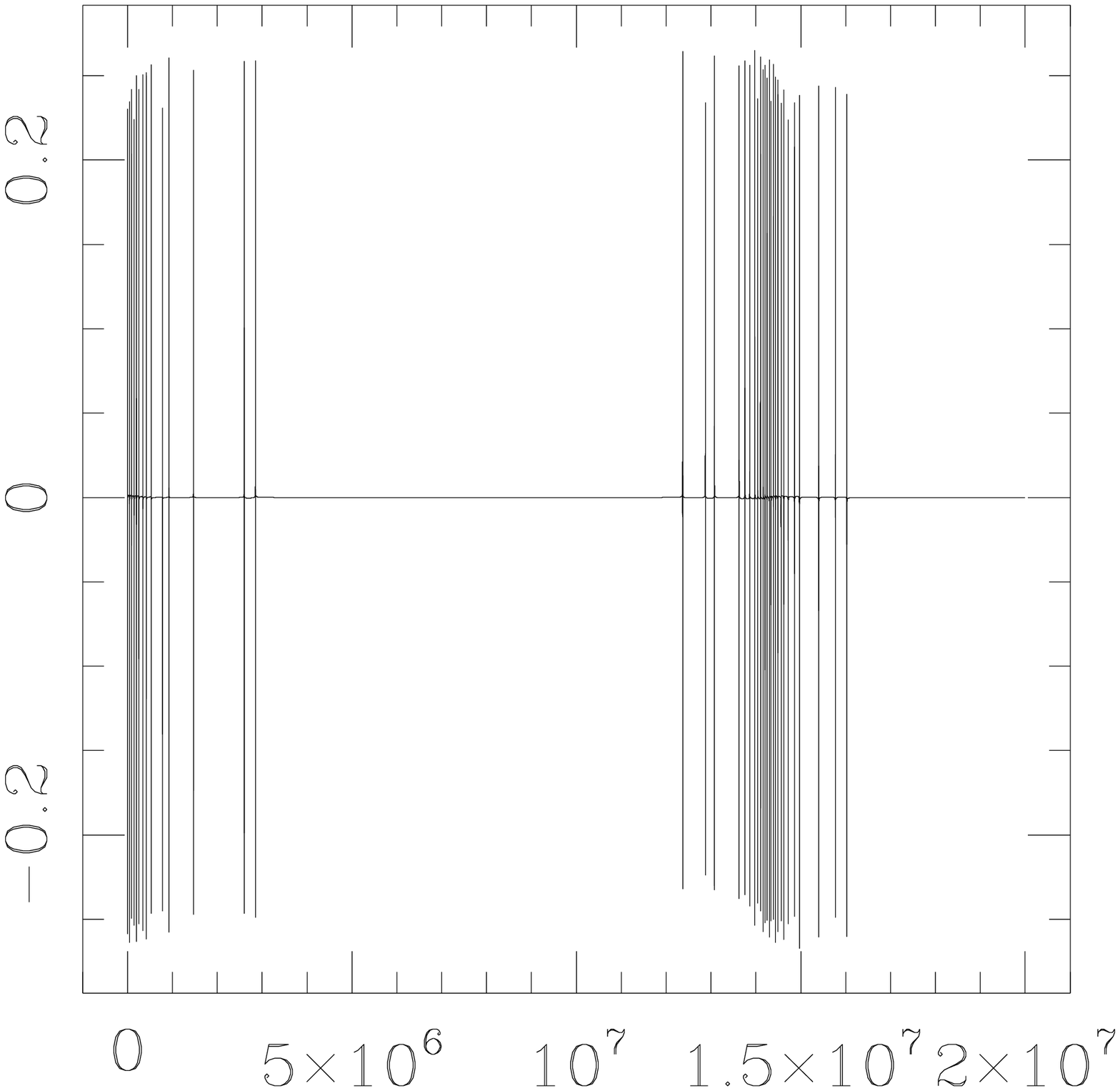,width=1.75in}
\psfig{file=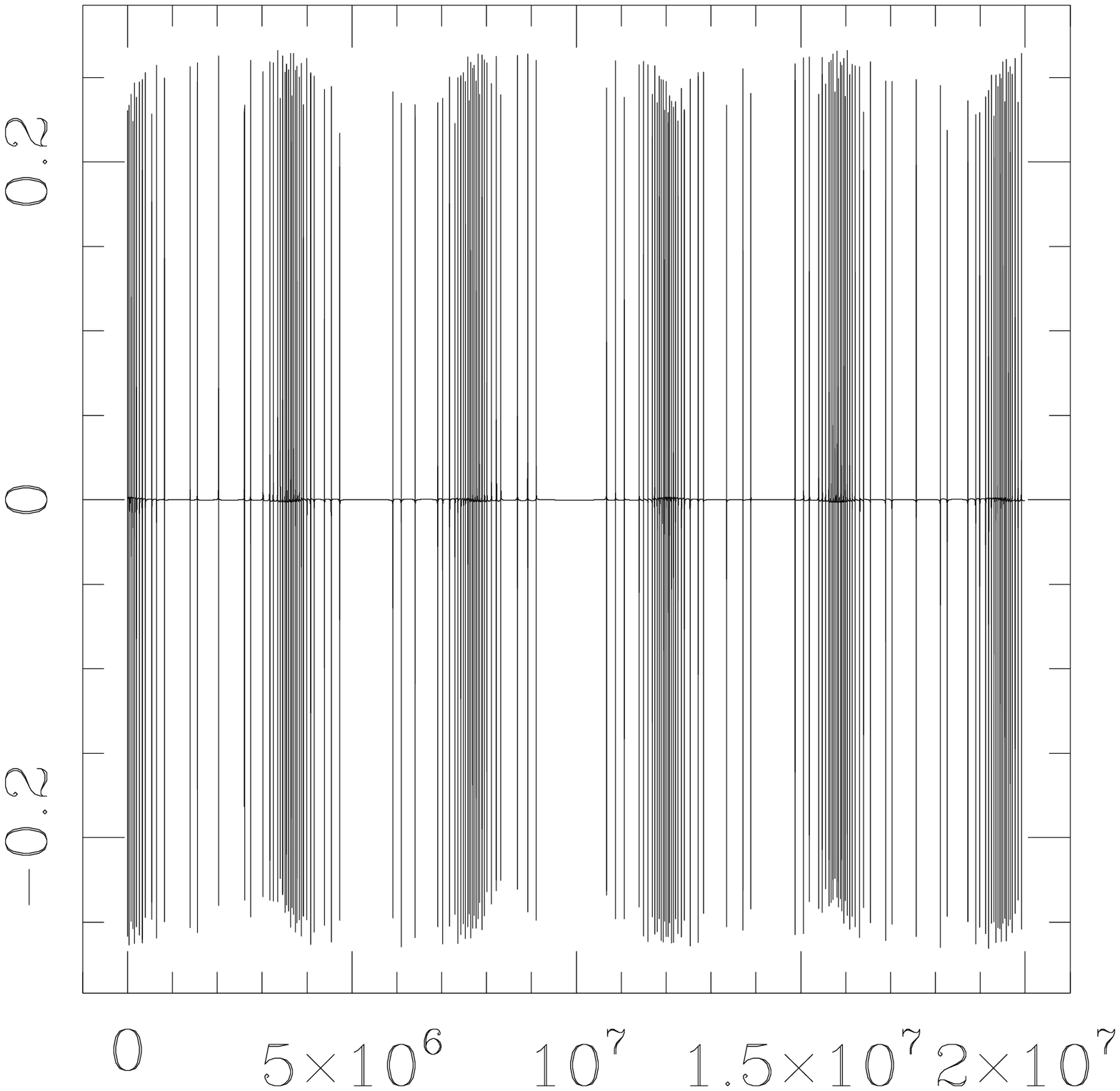,width=1.75in}}
\caption{The waveform $h_+$ for pairs selected from the initial
conditions in fig.\ \ref{fbb}.
Both orbits begin with $\theta_1=10{}^o$.  
The left panel began with
$\theta_2=128{}^o$ while the right panel began with
$\theta_2=131{}^o$.  
The extreme angles were randomly
chosen from the fractal set for illustration.  Chaos is seen with more
temperate angles as in fig.\ \ref{orb4}.
\label{wave}}  \end{figure}

Compact pairs with 
initial conditions drawn from near the fractal basin boundaries will
result in unpredictable outcomes.  They will have correspondingly
unpredictable waveforms.  The waveforms for pairs selected from the initial
conditions in fig.\ \ref{fbb} are shown in fig.\ \ref{wave}.
The orbits begin with nearly identical initial conditions.
Although the difference in
initial angles is only $3{}^o$, the waveforms are entirely
different.
The first pair separates while the second pair executes many thousands of
orbits.

It should be emphasized that orbits within smooth basins can still be
chaotic.  Well within the white stable basins, many orbits will precess
chaotically as does the orbit of fig.\ \ref{orb4}.  Similarly, many of
the escape orbits and the merger orbits will chaotically scatter
before reaching their final outcome.  Fractal basin boundaries are a
fairly blunt tool, insensitive to some manifestations of chaos.
Therefore while fractal basin boundaries do prove the dynamics is
chaotic, smooth basins are inconclusive.

With the radiative reaction included, the pair goes from
an energy conserving scattering system to a dissipative one.  
In any stability analysis, dissipation must be turned off to distinguish 
instability to the onset of chaos from instability to merger from simple 
energy loss.  Once the chaos has been identified, radiative back reaction
can readily be incorporated and we do so now.
Under the effects of dissipation, some orbits will sweep through the chaotic
region of phase space as they inspiral.
The surface of section is not useful for a dissipative
system since the radius of the orbit is shrinking as energy is lost to gravity
waves.
However, fractal basin boundaries are still effective 
at identifying extreme sensitivity to initial conditions.
Another advantage is that several thousand orbits can be scanned at once.
We use this method to show
that dissipation does not obliterate the chaos. 

\begin{figure}
\centerline{\psfig{file=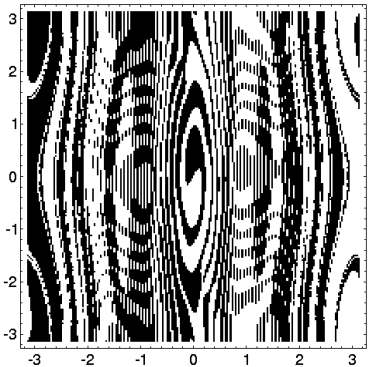,width=2.in}}
\caption{The fractal basin boundaries with
dissipation 
included.
The parameters are $m_2/m_1=1.4/10$ and spins 
$S_1=m_1^2$ and $S_2=0.7\ m_2^2$.  The orbits begin with $x_i/m=26,\dot
y_i=0.15$ and $z_i=0$.  The pair can execute
anywhere from $0$ to ${\cal O}(40)$ orbits before coalescence.
The initial angles $(\theta_1,\theta_2)$ are varied from $-\pi$ to $\pi$.
$300\times 300$ orbits shown.  
\label{strange}}  \end{figure}

As energy is lost the 
binary pairs tend to coalesce
in such a way that $r\rightarrow $merger 
is an attractor in phase space that can be
described by another fractal set.  To show this, we again look at
an initial condition slice through phase space.  We evolve each of these pairs
under the influence of the radiative reaction force.  We need to color code
the initial conditions on the basis of some well defined outcome.
Since all pairs
considered coalesce, we have to select some other criterion than that used
above.  We choose to color code 
the initial
location in the $(\theta_1,\theta_2)$ plane white if the
pair approach merger from below the $z$-axis and black if they approach 
merger from
above the $z$-axis.  The resultant
fractal is shown in fig.\ \ref{strange}.  Another criterion could have been
selected and in this sense the basin boundaries are crude, as already mentioned,
but they are nonetheless powerful at signaling the presence of chaos.
The conclusion to draw from this figure
 is that there is extreme
sensitivity to initial spin angles for 
rapidly spinning, 
inspiralling $10M_\odot $ black hole and $1.4 M_\odot $ neutron 
star binaries.  The pairs will inspiral along different paths as a
result of this sensitivity and therefore will have disparate
waveforms.
Similar chaotic sets have also been found for different binary mass
ratios and orbital parameters.

This work demonstrates the existence of chaotic
regions of phase space.  At least some orbits will move into this
chaotic region as they inspiral.  Of course some orbits will 
still be regular such as circular inspiral with 
spins exactly aligned with the orbital angular momentum.
A systematic scan of all parameters is needed to ascertain when the
dynamics is predictable and regular and when it is chaotic.
A quantitative comparison of the waveforms from a chaotic orbit
against a circular template is also needed to evaluate how seriously
chaos would deter detection.  Given that eccentricity in an otherwise
simple orbit can greatly diminish the signal when matched against a
circular template \cite{marp}, the 
chaotic precession does not bode well.
Still, the luminosity in gravity waves
is enhanced for some of these wilder orbits \cite{cf}, 
as was already seen along
the regular homoclinic orbits \cite{us}.  
Though unlikely,
an optimist might hope that direct detection of these gravity waves
will be possible if the signal is boosted substantially above the
noise, relieving the dependence on a theoretical template.

The inherent difficulty in the direct detection of gravity waves
highlights the importance of indirect methods of detection.
Corroborating evidence for gravity waves in
electromagnetic observations may be promising.  
Chaos can have
unexpected benefits if the black hole is able to 
capture the light from a luminous companion for many chaotic orbits
before some of the light escapes.
Such chaotic scattering of a
pulsar beam around a central black hole could lead to a diffuse glow
around the pair \cite{me}.  While this signature is likely to be
faint, any confirmation of a gravity wave signal will be welcome.

\vskip 15truept

I am grateful to E.J.Copeland, R.O'Reilly, and N.J.Cornish for 
their valuable input.
This work is supported by a PPARC Advanced Fellowship.

\end{document}